# High-contrast imaging of 180° ferroelectric domains by optical microscopy using ferroelectric liquid crystals

Guillaume. F. Nataf,[1,a)] Mael Guennou,[2,3] Giusy Scalia,[2] Xavier Moya,[1] Tim D. Wilkinson,[4] Jan P. F. Lagerwall[2,a)]

## AFFILIATIONS

[1] Department of Materials Science, University of Cambridge, 27 Charles Babbage Road, Cambridge, CB3 0FS, United Kingdom
[2] Department of Physics and Materials Science, University of Luxembourg, 162a Avenue de la Faïencerie, 1511 Luxembourg, Luxembourg
[3] Materials Research and Technology department, Luxembourg Institute of Sciences and Technology, 41 rue du Brill, 4422 Belvaux, Luxembourg
[4] Centre of Molecular Materials for Photonics and Electronics, Electrical Engineering Division, Department of Engineering, University of Cambridge, 9 JJ Thomson Avenue, Cambridge CB3 0FA, United Kingdom

[a)] Authors to whom correspondence should be addressed: gn283@cam.ac.uk, jan.lagerwall@lcsoftmatter.com

## ABSTRACT

Ferroelectric liquid crystals (FLCs) couple the direction of their spontaneous electric polarization to the direction of tilt of their optic axis. Consequently, reversal of the electric polarization by an electric field gives rise to an immediate and lasting optical response when an appropriately aligned FLC is observed between crossed polarizers, with one field direction yielding a dark image, and the opposite direction yielding a bright image. Here this peculiar electro-optic response is used to image, with high optical contrast, 180° ferroelectric domains in a crystalline substrate of magnesium-doped lithium niobate. The lithium niobate substrate contains a few domains with upwards electric polarization surrounded by regions with downward electric polarization. In contrast to a reference non-chiral liquid crystal that is unable to show ferroelectric behavior due to its high symmetry, the FLC, which is used as a thin film confined between the lithium niobate substrate and an inert aligning substrate, reveals ferroelectric domains as well as their boundaries, with strong black and white contrast. The results show that FLCs can be used for non-destructive read-out of domains in underlying ferroelectrics, with potential applications in e.g. photonic devices and non-volatile ferroelectric memories.

Over the past decades, the focus of research on ferroelectricity has shifted from bulk materials to thin films and nanostructures, where interfaces are key.[1–3] Interfaces in ferroelectric materials are generally characterized by bound charges resulting from discontinuities in electric polarization **P**, which can generate stray fields.[4] These charges, and/or the free charges screening them, modify the surface potential of bulk domains[5–7] and change interface reactivity.[8–10] Among other applications,[4] the interfacial charges open a pathway to ferroelectric domain imaging[11–17] and chemical control of electric polarization.[18–21]

The ability to image ferroelectric domains, in such a way that opposite directions of electric polarization can be distinguished, with a damage-free, fast and easy-to-handle method, is critical for applications relying on domain engineering, such as photonic devices,[22–26] where the quality of patterned 180° ferroelectric domains has to be assessed. If the method allows continuous monitoring of the electric polarization direction, it could also be a powerful solution for non-destructive read-out of ferroelectric memories.[2] Several techniques have been shown to be relevant for 180° domains imaging.[27] However, they still fail to address all challenges at once. For example, techniques with high spatial resolutions are too slow for large scale millimetric imaging (e.g. piezoresponse force microscopy[28] and transmission electron microscopy) or difficult to perform on insulating materials (e.g. photoemission electron microscopy[29]). Optical methods are ideally suited for fast large-scale imaging but second-harmonic generation[30] requires complex and expensive experimental setups, and polarized light microscopy can only resolve 180° domain walls (not 180° domains).[31] This leaves domain detection by surface modifications as an alternative, but etching is destructive[32] and powder decorated crystals are difficult to handle.[11] Decoration by liquid crystals (LCs) has also been suggested. It has been reported that nematic LCs can be used to distinguish between 180° domains in several ferroelectrics, using polarized light microscopy.[14–16,33–38] Moreover, measurements as a function of electric field or temperature have been successful in imaging domain dynamics. Most studies have been performed close to room temperature using nematic

LC N-(4-Methoxybenzylidene)-4-butylaniline (MBBA), or a mixture of MBBA and N-(4-Ethoxybenzylidene)-4-butylaniline (EBBA), but also at other temperatures with alkylphenyl-cyclohexylbenzoate and mixtures of cyanobiphenyls. However, the mechanism leading to the observed contrast remains unclear, as explained in detail in Ref. 1.

In the original report,[16] it was suggested that dipole moments of the LC MBBA align in opposite directions on neighboring domains. However, given the $D_{\infty h}$ symmetry of the nematic (N) phase, there cannot be sensitivity of the orientation of the *director* **n** (identifying the preferred orientation of the main molecule symmetry axis, and equivalent to the optic axis direction) to the *direction* of an applied electric field. Thus, the change in electric polarization direction between adjacent domains cannot on its own give rise to a contrast in a polarized light microscope. A few years later, Glogarová *et al.* suggested that the observed contrast resulted from an ionic interaction between the molecules of MBBA and $NH_3$ groups at the surface of triglycine sulphate (TGS),[14,33] and Nakatani *et al.* suggested that instead it results from different absolute values of the electric polarization in different domains.[34] This second approach would be consistent with the response of nematic LCs to the remanent polarization of imperfectly poled ferroelectric thin films.[39,40] If these were the mechanisms, it would be unlikely that MBBA, and nematic LCs in general, could be used to image 180° domains in a large number of ferroelectric materials.

Here, we describe a polarization direction-mediated alignment of the optic axis in surface-stabilized ferroelectric liquid crystals[41] (SSFLCs), clearly differentiating between domains with opposite polarizations through distinctly different optical behavior on $\mathbf{P}_{up}$ and $\mathbf{P}_{down}$ domains, respectively. A 5% magnesium-doped lithium niobate (Mg-$LiNbO_3$) single crystal from PI-KEM Ltd was chosen as the bulk ferroelectric substrate, initially uniformly poled with a single direction of polarization orthogonal to the substrate across the whole substrate. A random pattern of 180° re-poled ferroelectric domains with hexagonal shapes was then created at room temperature by applying high voltage pulses through liquid electrodes.[42]

The single crystal was annealed at 200°C for 30 minutes to erase the internal field[43] and immersed successively in acetone, isopropanol and distilled water for 10 minutes each time, in order to clean the surface. This single crystal formed one substrate of the LC sample cell, the other being a standard polyimide-coated glass substrate from a commercial LC cell (EHC). The polyimide promotes planar alignment of the director and it had been unidirectionally rubbed in order to define the preferred orientation in the plane of the substrate of the LC director in the nematic and smectic SmA/SmA* (the asterisk signifying chirality) phases.

As FLC, we chose a room temperature multi-component mixture W-235B, purchased from the group of Prof. R. Dabrowski, Military University of Warsaw (Poland), exhibiting a long natural pitch, facilitating efficient surface stabilization, and a birefringence in the range 0.12–0.14. We used a minimum amount of the FLC, confining it between the $LiNbO_3$ substrate and the polyimide-coated glass substrate without any spacers, pressing the substrates together by hand while the LC was heated up to the isotropic (I) phase to obtain a very small but uncontrolled cell gap (estimated on the order of 1 µm). The substrates are kept together by capillary forces from the FLC. The W-235B mixture exhibits the phase sequence "Crystal -10 SmC* 86 SmA* 90 N* 106 I" /°C (temperature indications from manufacturer, ignoring two-phase coexistence). For comparison, we chose a non-chiral LC, 2-(4-hexyloxyphenyl)-5-octylpyrimidine (6OPhPy8, purchased from Synthon, Germany), which exhibits the phase sequence "Crystal 28 SmC 45 SmA 58 N 65 I" /°C.

Textures were studied using an Olympus BX51 polarized light microscope working in transmission with a U-25LBD daylight color filter inserted. This compensates for the reddish illumination from the microscope lamp but instead leaves a light bluish tone to images devoid of strong colors. The temperature of the sample was controlled with a Linkam LTS120 hot stage. The LCs were heated up to the I phase and then cooled down towards room temperature at cooling rate ranging between 5°C $min^{-1}$ and 30°C $min^{-1}$, taking texture photographs in the process.

SSFLCs arise by confining a chiral smectic-C (SmC*) LC between tightly spaced flat substrates that impose a planar alignment, forcing the smectic layers to orient (largely) orthogonal to the cell [Figs. 1(a) and 1(b)] and unwinding the helix that spontaneously develops in bulk SmC* materials. In the SmC* phase (as well as in its non-chiral analog SmC), the director **n** is tilted by an angle θ away from the smectic layer normal **k**, effectively defining a tilt cone [Fig. 1(c)], since any *direction* of tilt is allowed in the absence of external forces. In contrast to nematics, the reduced symmetry of the SmC* phase gives it a polar, thus direction-sensitive, response to an electric field applied along the smectic layers, because it exhibits a non-zero spontaneous electric polarization $\mathbf{P}_s$ oriented perpendicular to **n** and **k** [Fig. 1(c)], $\mathbf{P}_s \propto \mathbf{n} \times \mathbf{k}$.[44] The unwinding of the SmC* helix is of key importance, as this yields the two-state ferroelectric behavior[41] in contrast to the effectively antiferroelectric response of bulk SmC* phases,[45] where the continuous rotation of $\mathbf{P}_s$ in the helix leads to a cancelation of the electric polarization on the scale of a helix period [Fig. 1(d)].

This geometrical coupling between **n** and $\mathbf{P}_s$ means that an electric field applied along the smectic layers, orthogonally to the substrate plane of our samples, forces **n** to tilt right or left, depending on the direction of the field and of the **n**–$\mathbf{P}_s$ coupling. The benefit is that **n** rotates 180° around the SmC* tilt cone, corresponding to a reorientation of the optic axis by twice the tilt angle, 2θ, if one changes the electric field direction between up and down. This inversion of electric field is what happens when one moves across a domain wall between 180° ferroelectric domains at the surface of a ferroelectric substrate. Thus, the electro-optic response of the SSFLC gives rise to a strong contrast between the 180° ferroelectric domains when the sample is observed in polarized light microscopy, as we demonstrate here using sandwiched SSFLC on a $LiNbO_3$ single crystal.

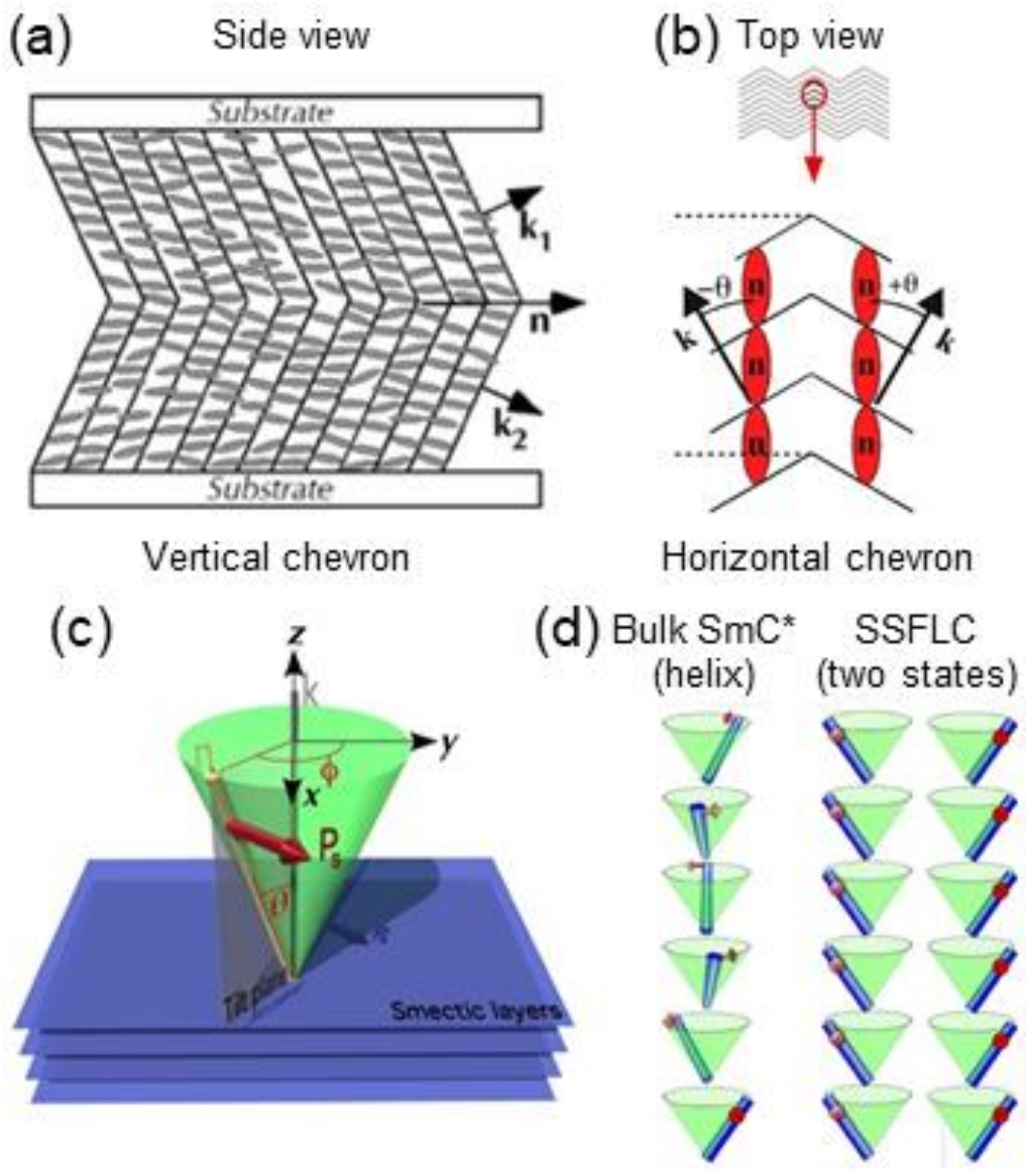

**FIG. 1.** An SSFLC sample is produced by filling a SmC*-forming LC between closely spaced flat substrates. In the SmC/SmC* phase, the layer contraction upon director tilt leads to formation of chevrons, which can be of (a) vertical or (b) horizontal type. (c) The reduced symmetry of a SmC* phase allows for the development of a spontaneous electric polarization $\mathbf{P}_s$, orientated perpendicular to the tilt plane, spanned by **n** and the layer normal **k** (reproduced with permission from ChemPhysChem 7, 20 (2006) © 2006 WILEY-VCH). (d) Without the confinement, the SmC* develops a helix that cancels out the electric polarization on the scale of a helix pitch. When the helix is suppressed by closely spaced substrates, only the two states are allowed where **n**, and thus the optic axis, lies in the substrate plane, tilted ±θ from **k**. This corresponds to $\mathbf{P}_s$ being directed up or down with respect to the substrate plane.

We observe the contrast evolution in the different phases of the FLC at different temperatures (I, N*, SmA*, SmC*), and compare with the evolution of a non-chiral reference LC, that thus does not exhibit ferroelectricity although it has the corresponding (non-chiral) phase sequence (I, N, SmA, SmC). The ferroelectricity of SSFLCs is a generic property of SmC* LCs in which the helix has been unwound by confinement, and the response is generic to applied electric fields, hence our findings apply to any ferroelectric substrate and can be realized with any SmC* LC with appropriate alignment and helix unwinding.

Fig. 2 shows the polarized light microscopy texture evolution on cooling of the FLC mixture W-235B [Figs. 2(a–d)] and the non-chiral reference compound 6OPhPy8 LC [Fig. 2(e–h)]. Images in the isotropic phase [Figs. 2(a) and (e)] are identical to images of $LiNbO_3$ obtained in air where domains can be localized only because their domain walls appear as shiny stripes, as a result of local strain-induced birefringence.[31,46] In the N*/N phase [Figs. 2(b) and (f)], no contrast is observed because the optic axis of the LC is along the rubbing direction of the counter substrate, regardless of $LiNbO_3$ domains. The contrast at domain walls is barely visible, as the birefringence of the LC now dominates the optical behavior, rendering the weak contribution from the strain-induced birefringence at domain walls in $LiNbO_3$ negligible. In the SmA*/SmA phase [Figs. 2(c) and (g)], the SSFLC already reveals slightly the domain structure [Fig. 2(c)] while domains are still not visible with 6OPhPy8 [Fig. 2(g)]. The reason the SmA* phase responds to the domain electric polarization is most likely the electroclinic effect,[44] also called soft mode, which yields a slight director tilt in response to the field, the effect being significant even a few degrees above the transition to the SmC* phase.[47]

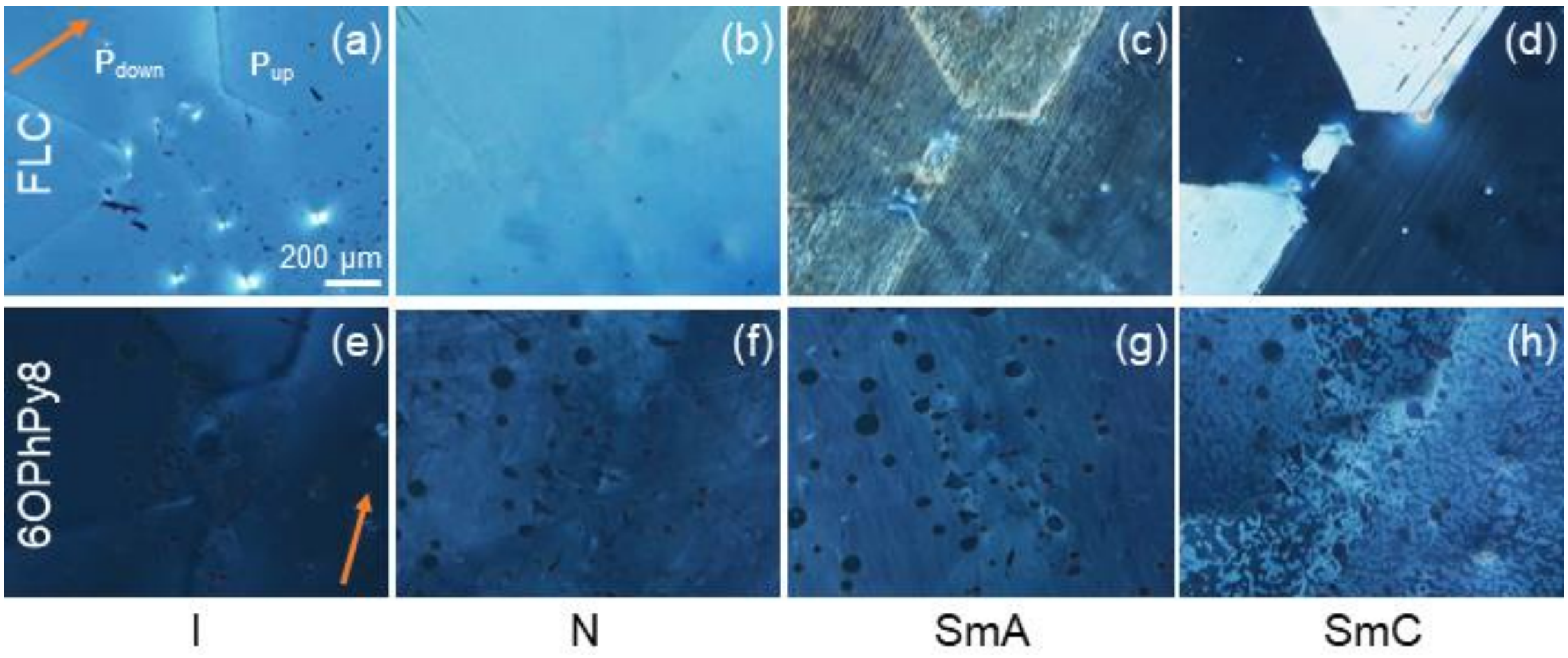

**FIG. 2.** Polarized light microscopy texture evolution on cooling of the (a-d) SSFLC mixture W-235B in the (a) I, (b) N*, (c) SmA* and (d) SmC* phases at 110°C, 95°C, 88°C and 25°C, respectively; and of (e-h) the non-chiral reference LC 6OPhPy8 in the (e) I, (f) N, (g) SmA and (h) SmC phases at 74°C, 61°C, 50°C and 31°C, respectively. Note that the sample is rotated somewhat between (a) and (b) and between (b) and (c), and that the sample with 6OPhPy8 contains air bubbles, appearing as black circles. In (a), $\mathbf{P}_{up}$ and $\mathbf{P}_{down}$ refer to the direction of the electric polarization in the domains of Mg-$LiNbO_3$. The arrows in (a) and (e) indicate the rubbing direction for the SSFLC mixture and 6OPhPy8, respectively. The bluish tint is due to the daylight filter used in the microscope. The camera was adjusting exposure automatically, hence brightness levels are not comparable between panels.

In the SmC* phase, where the confined W-235B is in the SSFLC state, there is a strong contrast with reverse-poled domains appearing white and the virgin background black, for the sample orientation in Fig. 2(d). The two permissible SSFLC states [(Fig. 1(d)] are distinguished by a reorientation of the optic axis by an angle 2θ, giving almost maximum brightness for the re-poled domains, as the sample has been rotated such that the virgin background has its optic axis along the polarizer of the microscope (thus appearing dark). The layer normal **k** is along the rubbing direction of the polyimide-coated substrate, as the layer orientation was defined on cooling from the N*–SmA* transition. The two domain types thus have their optic axis oriented at ±θ with respect to the rubbing direction.

In the SmC phase of 6OPhPy8 the optic axis also tilts away from **k** by ±θ, but there is no spontaneous electric polarization that couples to the tilt. This means that the opposite electric fields arising from the 180° domains of $LiNbO_3$ have no impact on whether the optic axis tilts clockwise or anticlockwise away from **k**, and we thus see a random spontaneous separation into small right- and left-tilting domains [Fig. 2h].

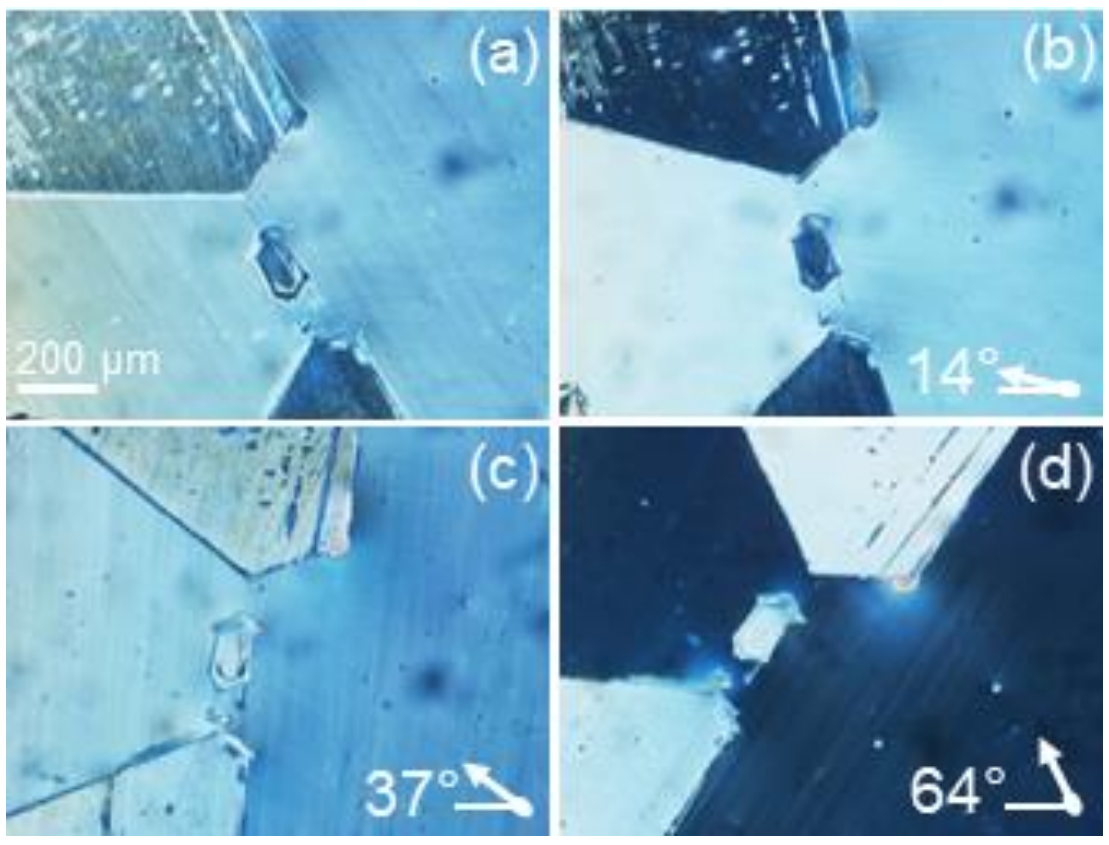

**FIG. 3.** Room temperature polarized light microscopy texture in the SmC* phase of the SSFLC, as a function of sample rotation, just after cooling from the I phase (via N* and SmA*). In (d), the sample has been rotated by 64° compared to (a).

In the SmC* phase of the FLC, rotation of the sample by 2θ = 50° (Movie 1 in the Supplementary Material) leads to contrast reversal between domains [(Figs. 3(b) and (d)] and domain walls [Figs. 3(a) and (c)]. This confirms that the electric polarization of the domains of $LiNbO_3$ (~70 μC $cm^{-2}$),[48] and the resulting stray electric field, controls the tilt direction of the SSFLC. Note that the domain walls are dark when both domains are bright [Fig. 3(c)], at a sample orientation with the layer normal **k** along the polarizer. The projection of **n** (and thus of the optic axis) into the sample plane is thus along **k** at the domain walls. As illustrated in Fig. 1c–d, this means that $\mathbf{P}_s$ is aligned in the sample plane, due to the geometrical coupling $\mathbf{P}_s \propto \mathbf{n} \times \mathbf{k}$. This suggests that the liquid crystal alignment is dictated by the substrate not only on top of domains, but that it even follows the reversal of the electric polarization taking place across the domain wall.[7]

On the time scale of the characterization experiments (~1 h), the optical contrast is stable, which indicates that it does not arise from an enhancement of the polarization in the domains on cooling through the pyroelectric effect. However, after leaving the sample for 15 hours at room temperature, the contrast is lost in these samples as the original texture breaks up into an irregular granular one that does not reflect the electric field arising from the poled domains of the ferroelectric substrate (Supplementary Material, Fig. S1). This may be related to ions in the FLC redistributing (slowly) to the interface with the $LiNbO_3$ substrate or to structural reorganization within these samples with mobile unglued substrates, as discussed in the Supplementary Material. Irrespective of the details of the mechanism, and considering that SSFLC displays with reliable long-term performance have been realized, we are confident that this can be resolved by proper control of the sample configuration and optimization of the LC mixture.

In summary, we demonstrated that SSFLCs accurately reveal the 180° electric polarization contrast of opposite domains in ferroelectrics, here exemplified by a Mg-$LiNbO_3$ single crystal, with high-contrast as viewed in a polarized light microscope. Contrary to previous reports of LC-enabled visualization of ferroelectric domains, which exclusively used nematic LCs, whose symmetry is too high to distinguish between opposite directions of electric field, the behavior here fully complies with the expected behavior of a surface-stabilized ferroelectric SmC* phase. The results show that SSFLCs can be used for non-destructive read-out of domains in ferroelectrics, of potential use, e.g., in photonic devices and non-volatile ferroelectric memories. Having established the physical foundations of the method, future work will include in-depth systematic investigations of its practical aspects. In particular, engineering efforts will be required to reach an optimal spatial resolution and extend the stability in time of the LC response to the polarization of the substrate. The spatial resolution is ultimately limited by the size of the domains in the FLCs. Based on literature and standard LC technology, we believe that a resolution of ~4 μm can be achieved,[49,50] which would be sufficient, for instance, for typical applications of $LiNbO_3$ in non-linear optics.[23,25]

## SUPPLEMENTARY MATERIAL

See supplementary material for videos of burst-like textural reconfiguration upon fast cooling and texture variation upon rotation of the SSFLC.

## ACKNOWLEDGEMENTS

GFN would like to thank the Royal Commission for the Exhibition of 1851 for the award of a Research Fellowship. X. M. is grateful for support from the Royal Society.

## DATA AVAILABILITY

The data that supports the findings of this study are available within the article and its supplementary material.